\documentclass[twocolumn,showpacs,floatfix,aps]{revtex4}
\usepackage{amssymb,amsmath}
\usepackage[dvips]{graphics}
\usepackage{epsfig}
\newcommand{\bea}{\begin{array}}
\newcommand{\ear}{\end{array}}

\newcommand{\bege}{\begin{equation}}
\newcommand{\enge}{\end{equation}}

\newcommand{\beq}{\begin{eqnarray}}\newcommand{\benu}{\begin{enumerate}}\newcommand{\enu}{\end{enumerate}}
\newcommand{\eeq}{\end{eqnarray}}

\newcommand{\me}{\frac{1}{2}}
\newcommand{\noi}{\noindent}

\newcommand{\lie}{\text{{\it\char'44}}}
\newcommand{\be}{\beta}

\begin{document}

\title{Variation in the luminosity of Kerr quasars due to extra dimension in brane Randall-Sundrum model}

\author{Rold\~ao da Rocha}
\email{roldao@ifi.unicamp.br}
\affiliation{IFGW, Universidade Estadual de Campinas,\\
CP 6165, 13083-970 Campinas, SP, Brazil.}

\author{Carlos H. Coimbra-Ara\'ujo}
\email{carlosc@ifi.unicamp.br}
\affiliation{Departamento de Astronomia, Universidade de S\~ao Paulo, 05508-900
S\~ao Paulo, SP, Brazil.\\and\\
IFGW, Universidade Estadual de Campinas,\\
CP 6165, 13083-970 Campinas, SP, Brazil.}


\begin{abstract}
We propose an alternative theoretical approach  showing how the existence of an extra dimension in Randall-Sundrum model can estimate 
the correction in the horizon of Schwarzschild and Kerr black holes, and consequently its measurability in terms of the variation of quasar luminosity,
which can be caused by a imprint of an extra dimension endowing 
the geometry of brane-world scenario in  AdS$_5$ bulk. The rotation effects cause a more prominent correction in Kerr horizon radii 
than in Schwarzschild (static black hole) radius, via brane-world effects, and the consequent bigger variation in the luminosity 
in Kerr black holes quasars. 
This paper is intended to investigate the variation of luminosity due to accretion of gas in Schwarzschild and Kerr black holes (BHs)
in the center of quasars, besides also investigating the variation of luminosity in supermassive BHs by brane-world effects, 
using Randall-Sundrum model. 
 
\end{abstract}
\bigskip\bigskip
\maketitle

\section{Introduction}

The possibility concerning the existence of extra dimensions is one of the most astonishing 
aspects of string theory and the formalism of $p$-branes.
In spite of this  possibility, extra dimensions still remain up to now unaccessible and obliterated to experiments.
An alternative approach to the compactification of extra dimensions, provided by, e.g., Kaluza-Klein (KK) and string theories \cite{gr,zwi,zwi1,zwi2}, 
involves an extra dimension which is not compactified, as pointed by, e.g.,  Randall-Sundrum (RS) model \cite{Randall1,Randall2}. 
This extra dimension implies
deviations on Newton's law of gravity at scales below about 0.1 mm, where objects may be indeed gravitating in more
dimensions. The electromagnetic, weak and strong forces, as well as all the matter
in the universe, would be trapped on a brane with three spatial dimensions, and only gravitons would be allowed to leave the surface
 and move into the full bulk, constituted by an AdS$_5$ spacetime, as prescribed by, e.g., 
 in RS model \cite{Randall1,Randall2}.

At low energies, gravity is localized on the brane and general relativity is recovered, but at high energies,
 significant changes are introduced in gravitational dynamics, forcing general relativity to break down
to be overcome by a quantum gravity theory \cite{rov}.
A plausible reason for the gravitational force appear to be so weak in relation to other forces can be its dilution in possibly existing extra dimensions
related to a bulk, where $p$-branes \cite{gr,zwi, zwi1, zwi2, Townsend} are embedded. $p$-branes 
are good candidates for brane-worlds \cite{ken} because they possess gauge symmetries \cite{zwi, zwi1, zwi2} and automatically incorporate a quantum theory of gravity. The
gauge symmetry arises from open strings, which can collide to form a closed string that can leak into the higher-dimensional
bulk. The simplest excitation modes of these closed strings
correspond precisely to gravitons.
An alternative scenario can be achieved by RS 
  model \cite{Randall1,Randall2}, 
which induces a volcano barrier-shaped effective potential for gravitons around the brane \cite{Likken}.
The corresponding spectrum of gravitational perturbations has a massless bound state on the brane,
and a continuum of bulk modes with suppressed couplings to brane fields. These bulk modes introduce small
corrections at short distances, and the introduction of more compact dimensions does not affect the localization of matter fields.
However, true localization takes place only for massless fields \cite{Gregory}, and in the massive case the bound state becomes metastable,
being able to leak into the extra space. This is shown to be exactly the case for astrophysical massive objects,
where highly energetic stars and the process of gravitational collapse, which can originate black holes, leads to deviations from the $4D$ general relativity problem.
There are other interesting and astonishing features concerning RS models, such as the AdS/CFT correspondence of a RS infinite AdS$_5$ brane-world, without matter 
fields on the brane, and 4$D$ general relativity coupled to conformal fields \cite{Randall1,Randall2,Maartens}. 

We precisely investigate the consequences of the deviation of a Schwarzschild-like term in a 5$D$ spacetime metric, predicted 
by RS1 model in the correction of the Schwarzschild radius of a BH, and also the deviation of a Kerr-like term, corresponding 
to a rotating BH. We show that, for fixed effective extra dimension size, supermassive BHs (SMBHs)  
give the upper limit of variation in luminosity of quasars, and although the method used holds for any other kind of BH, such as mini-BHs and 
stellar-mass ones, we shall use SMBHs parameters, where the effects are seen to be more notorious.
 It is also analyzed how the quasar luminosity variation behaves as a function of the AdS$_5$ bulk radius 
$\ell$, for various values of BH masses, from $10$ to $10^6$ solar masses. 

The search for observational evidence of higher-dimensional gravity is an
important way to test the ideas that have being come from string theory. This
evidence could be observed in particle accelerators or gravitational wave
detectors. The wave-form of gravitational waves produced by black holes, for
example, could carry an observational signature of extra dimensions, because
brane-world models introduce small corrections to the field equations at high
energies. But the observation of gravitational waves faces severe limitations
in the technological precision required for detection. This is a undeniable fact.
Possibly, an easier manner of testing extra dimensions can be via the
observation of signatures in the luminous spectrum of quasars and microquasars. This is the
goal of this paper, which is the first of a series of papers we shall present. Here we show the possibility of detecting brane-world
corrections for big quasars associated with Schwarzschild and Kerr SMBHs by their luminosity observation. In the next article we shall see that 
these corrections are more notorious in mini-BHs, where the Reissner-Nordstr\o m radius in a brane-world scenario 
shall be shown to be $10^4$ times bigger than standard Reissner-Nordstr\o m radius associated with mini-BHs. 
Indeed, mini-BHs are shown to be much more  
sensitive to brane-world effects.
In the last article of this series we also present an alternative 
possibility to detect electromagnetic KK modes due to perturbations in black strings \cite{ma,soda}.

This article  is organized as follows: in Section 2 after presenting Einstein equations
in AdS$_5$ bulk and discussing the relationship between the electric part of Weyl tensor 
and KK modes in RS1 model, the deviation in Newton's 4$D$ gravitational potential
is introduced in order to predict the deviation in Schwarzschild form and 
its consequences on the variation in quasar luminosity. For a static spherical metric 
on the brane the propagating effect of 5$D$ gravity is shown to 
arise only in the fourth order expansion in terms of the Taylor's of the normal coordinate out of the brane. 
In Section 3 the variation in quasar luminosity is carefully investigated, by finding 
the correction respectively  in the Schwarzschild and in the Kerr (external) horizon, caused by brane-world effects. 
In Appendix we carefully analyze twenty six \emph{ansatzen} for the correction to the deviation in Kerr form, by brane-world effects. 
All results are illustrated by graphics 
and figures.

\section{Black holes on the brane}
In a brane-world scenario given by a 3-brane embedded in an AdS$_5$ bulk the Einstein field equations read 
\begin{eqnarray}\label{123}
&&G_{\mu\nu} = -\frac{1}{2}{\Lambda}_5g_{\mu\nu}\nonumber\\
&&+ \frac{1}{4}\kappa_5^4\left[TT_{\mu\nu} - T^\alpha _{ \nu}T_{\mu \alpha} + \frac{1}{2}g_{\mu\nu}(T^2 - T_{\alpha\beta}^{\;\;\;\;\alpha\beta})\right] - E_{\mu\nu},\nonumber
\end{eqnarray}
\noindent where $T = T_\alpha^{\;\;\alpha}$ denotes the trace of the momentum-energy tensor $T_{\mu\nu}$, $\Lambda_5$ denotes 
the 5$D$ cosmological AdS$_5$ bulk constant, and  $E_{\mu\nu}$ denotes the `electric' components of the Weyl tensor, that can be expressed by means of the
extrinsic curvature components $K_{\mu\nu} = -\frac{1}{2} \lie_n g_{\mu\nu}$ by \cite{soda}
\begin{equation}
E_{\mu\nu} = \lie_n K_{\mu\nu} + K_{\mu}^{\;\;\alpha}K_{\alpha\nu} - \frac{1}{\ell^2}g_{\mu\nu}
\end{equation}\noindent where $\ell$ denotes the AdS$_5$ bulk curvature radius. It corresponds equivalently 
to the effective size of the extra dimension probed by a 5$D$ graviton \cite{Likken, Randall1,Randall2,Maartens}
The constant  $\kappa_5 = 8\pi G_5$, where 
$G_5$ denotes the 5$D$ Newton gravitational constant, that can be related to the
4$D$ gravitational constant $G$ by $G_5 = G\ell_{\rm Planck}$, where $\ell_{\rm Planck} = \sqrt{G\hbar/c^3}$ is the Planck length.

As indicated in \cite{Randall1,Maartens}, ``table-top tests of Newton's law currently find no deviations down to the order
of 0.1 mm'', so that $\ell \lesssim $ 0.1 mm. Empar\'an et al \cite{emparan} provides a more accurate magnitude limit improvement on the AdS$_5$ curvature $\ell$, 
by analyzing the existence of stellar-mass BHs
on long time scales and of BH X-ray binaries. In this paper we relax the stringency $\ell \lesssim 0.01$ mm to the former table-top limit
$\ell \lesssim $ 0.1 mm. 

The Weyl `electric' term $E_{\mu\nu}$  carries an imprint of high-energy effects sourcing KK modes. It means that highly energetic
stars and the process of gravitational collapse, and naturally  BHs, lead to deviations from the 4$D$ general
relativity problem. This occurs basically because the gravitational collapse unavoidably produces energies high enough to make
these corrections significant. From the brane-observer viewpoint, the KK corrections in $E_{\mu\nu}$ are nonlocal, since they
incorporate 5$D$ gravity wave modes. These nonlocal corrections cannot be determined purely from data on the brane \cite{Maartens}.
The component $E_{\mu\nu}$ also carries information about the collapse process of BHs.
In the perturbative analysis of Randall-Sundrum (RS) positive tension 3-brane, KK modes consist of a continuous spectrum without any gap. It
generates a correction in the gravitational potential $V(r) =\frac{GM}{c^2r}$ to 4$D$ gravity at low energies from extra-dimensional effects \cite{Maartens}, 
which is
given by \cite{Randall1,Randall2}
\begin{equation}\label{potential}
V(r) = \frac{GM}{c^2r}\left[1 + \frac{2\ell^2}{3r^2} + \mathcal{O}\left(\frac{\ell}{r}\right)^4\right].
\end{equation}
\noindent
The KK modes that generate this correction are responsible for a nonzero $E_{\mu\nu}$. This term carries the modification to the weak-field field equations, as we have
already seen.
 The Gaussian coordinate $y$ denotes hereon the direction normal out of the brane into the AdS$_5$ bulk, in each point of the 3-brane\footnote{In general 
the vector field cannot be globally defined on the brane, and it is only possible if the 3-brane is considered to be parallelizable.}.

\begin{figure}
\includegraphics[width=8.7cm]{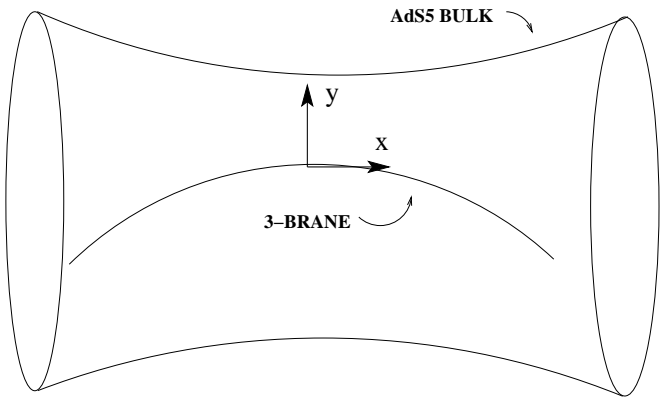}
  \caption{\small Schematic diagram of a slice of a 3-brane embedded in an AdS$_5$ bulk. The Gaussian coordinate
$y$ is normal to the brane and $x$ denotes spacetime coordinates in the brane.}
\label{fig:ads5}
\end{figure}

The RS metric is in general expressed as
\begin{equation}
^{(5)}ds^2 = e^{-2k|y|}g_{\mu\nu}dx^{\mu}dx^{\nu} + dy^2,
\end{equation}
\noindent
where $k^2 = 3/(2\ell^2)$,  
and the term $e^{-2k|y|}$ is called \emph{the warp factor} \cite{Randall1,Randall2,Maartens}, which
reflects the confinement role of the bulk cosmological constant $\Lambda_5$, preventing gravity from leaking
into the extra dimension at low energies \cite{Maartens,Randall1,Randall2}. 
The term $|y|$ clearly provides the $\mathbb{Z}_2$-symmetry
of the 3-brane at $y=0$. 

Concerning the anti-de Sitter (AdS$_5$) bulk, the cosmological constant can be written as $\Lambda_5 = -6/\ell^2$ and the brane is localized
at $y = 0$, where the metric recovers the usual aspect. 
The contribution of the bulk on the brane can be shown only to be due to the Einstein tensor, and can be expressed as
$\nabla_\nu G^{\mu\nu} = 0$, which implies that $\nabla_\nu (E^{\mu\nu} - S^{\mu\nu}) = 0$ \cite{Shiromizu},
where
\begin{equation}
S_{\mu\nu} := \frac{1}{4}\kappa_5^4\left[TT_{\mu\nu} - T^{\;\alpha} _{ \nu}T_{\mu \alpha} + \frac{1}{2}g_{\mu\nu}(T^2 - T_{\alpha\beta}^{\;\;\;\;\alpha\beta})\right]
\end{equation}
A vacuum on the brane, where $T_{\mu\nu} = 0$ outside a BH, implies that
\begin{equation}\label{21}
\nabla_\nu E^{\mu\nu} = 0.
\end{equation}
\noindent
Eqs.(\ref{21}) are referred to the nonlocal conservation equations. Other useful equations for the BH case are
\begin{equation}\label{ricci2}
G_{\mu\nu} = - \frac{1}{2}\Lambda_ 5g_{\mu\nu} - E_{\mu\nu}, \quad R = R^{\mu}_{\;\; \mu} = 0 = E^{\mu}_{\;\; \mu}.
\end{equation}
\noindent
 Therefore, a particular manner to express the vacuum field equations in the brane given by eq.(\ref{ricci2}) is
$E_{\mu\nu} = - R_{\mu\nu},$ 
where the bulk cosmological constant is incorporated to the warp factor in the metric.
One can use a Taylor expansion in order to probe properties of a static BH on the brane \cite{Da}, and for a vacuum brane metric,
we have, up to terms of order ${\mathcal O}(y^5)$ on, the following expression
\begin{eqnarray}\label{metrica}
&&\negthickspace g_{\mu\nu}(x,y) = g_{\mu\nu}(x,0) - E_{\mu\nu}(x,0)y^2 - \frac{2}{\ell}E_{\mu\nu}(x,0)|y|^3\nonumber\\
&&\negthickspace+\frac{1}{12}\left[\left({\Box} - \frac{32}{\ell^2}\right)E_{\mu\nu} + 2R_{\mu\alpha\nu\beta}E^{\alpha\beta} + 6E_{\mu}^{\; \alpha}E_{\alpha\nu}\right]_{y=0}\negthickspace y^4
\nonumber\end{eqnarray}
\noindent where $\Box$ denotes the usual d'Alembertian.
It shows in particular that the propagating effect of $5D$ gravity arises only at the fourth order of the expansion. For a static spherical metric on the brane
given by \begin{equation}\label{124}
g_{\mu\nu}dx^{\mu}dx^{\nu} = - F(r)dt^2 + \frac{dr^2}{H(r)} + r^2d\Omega^2,
\end{equation}
\noindent where $d\Omega^2$ denotes the spherical 3-volume element related to the geometry of the 3-brane,
 the projected electric component Weyl term on the brane is given by the expressions
\begin{eqnarray}
E_{00} &=& \frac{F}{r}\left(H' - \frac{1 - H}{r}\right),\;
E_{rr} = -\frac{1}{rH}\left(\frac{F'}{F} - \frac{1 - H}{r}\right),\nonumber\\
 E_{\theta\theta} &=& -1 + H +\frac{r}{2}H\left(\frac{F'}{F} + \frac{H'}{H}\right).
\end{eqnarray}
\noindent Note that in eq.(\ref{124}) the metric is led to the Schwarzschild one, if $F(r)$ equals $H(r)$.
The exact determination of these radial functions remains an open problem in BH theory on the brane \cite{Maartens,
rs05,rs06,rs07,rs08,rs09}.

These components allow one to evaluate the metric coefficients in eq.(\ref{metrica}). The area of the
$5D$ horizon is determined by $g_{\theta\theta}$. Defining $\psi(r)$ as the deviation from a Schwarzschild form for $H(r)$ \cite{Maartens,rs05,rs06,rs07,rs01,rs02,rs03,Gian}
\begin{equation}\label{h}
H(r) = 1 - \frac{2GM}{c^2r} + \psi(r),
\end{equation}
\noindent
where $M$ is constant, yields
\begin{eqnarray}\label{gtheta}
g_{\theta\theta}(r,y) &=& r^2  - \psi'\left(1 + \frac{2}{\ell}|y|\right)y^2\nonumber\\ +\negthickspace \negthickspace &&\negthickspace\negthickspace\left[\psi' + \frac{1}{2}(1 + \psi')(r\psi' - \psi)'\right]\frac{y^4}{6r^2} + \cdots
\end{eqnarray}
\noindent
It can be shown $\psi$ and its derivatives determine the change in the area of the horizon along the extra dimension \cite{Maartens}.
 For a large BH, with horizon scale $r \gg \ell$, it follows from eq.(\ref{potential}) that
\begin{equation}\label{psi}
\psi(r) \approx -\frac{4GM\ell^2}{3c^2r^3}.
\end{equation}
\noindent The formula above, together with eq.(\ref{potential}), can be directly 
deduced  from Randall-Sundrum analisys concerning small gravitational fluctuations in terms of KK modes, where
a curved background can support a  bound state  of the higher-dimensional graviton, which is localized in extra dimensions
\cite{Randall1,Randall2}.  Having found the KK spectrum of the effective 4$D$ theory, Randall \& Sundrum compute the 
non-relativistic gravitational potential between two particles of mass $m_1$ and $m_2$ on the 3-brane, 
that is the static potential generated by exchange of the zero-mode and continuum Kaluza-Klein mode propagators. The potential 
can be written as 
\begin{equation}
V(r) = G\frac{m_1m_2}{r} + \int_0^\infty \frac{Gm_1m_2\exp(-mr)\,dm}{k^2r}
\end{equation}\noindent which can be lead to eq.(\ref{potential}) and, consequently, to eq.(\ref{psi}).

\section{Variation in the luminosity of quasars and AdS curvature radius}
In this section we shall consider two cases:  quasars respectively formed by a Schwarzschild SMBH and by a Kerr SMBH.
\subsection{Corrections in the radius of Schwarzschild SMBHs by brane effects}
The observation of quasars (QSOs) in X-ray band can constrain the measure of the AdS$_5$ bulk curvature radius $\ell$, and indicate
 how the bulk is curled, from its geometrical and topological features.
QSOs are astrophysical objects that can be found at large astronomical distances (redshifts $z > 1$).
For a \emph{gedanken} experiment involving a static BH being accreted, in a simple model, the accretion efficiency $\eta$ is given by
\begin{equation}\label{eta}
\eta = \frac{GM}{6c^2R_{{\rm Sbrane}}},
\end{equation}
\noindent
where $R_{{\rm Sbrane}}$ is the Schwarzschild radius corrected for the case of brane-world effects. 
The luminosity $L$ due to accretion in a BH, that generates a quasar,
 is given by
\begin{equation}\label{dl}
L(\ell) = \eta(\ell) \dot{M}c^2,
\end{equation}
\noindent
where $\dot{M}$ denotes the accretion rate and depends on some specific model of accretion.

In order to estimate $R_{{\rm Sbrane}}$, fix $H(r) = 0$ in  eq.(\ref{h}), resulting in
\begin{equation}
1 - \frac{2GM}{c^2R_{{\rm Sbrane}}} - \frac{4GM\ell^2}{3c^2R_{{\rm Sbrane}}^3} = 0.
\end{equation}
\noindent This equation can be rewritten as 
\begin{equation}
R_{{\rm Sbrane}}^3 - \frac{2GM}{c^2}R_{{\rm Sbrane}}^2 - \frac{4GM\ell^2}{3c^2} = 0.
\end{equation}
\noindent Using Cardano's formula \cite{card} it follows that $R_{{\rm Sbrane}}$ can be exactly calculated as 
\begin{equation}\label{111}
R_{{\rm Sbrane}} = (a + \sqrt{b})^{1/3} + (a - \sqrt{b})^{1/3} + \frac{2GM}{3c^2},\end{equation}
\noindent where 
\begin{eqnarray}
a &=& \frac{2GM}{3c^2}\left(\ell^2 + \frac{4G^2M^2}{9c^4}\right),\label{rs3}\\
 b&=&\frac{4G^2M^2\ell^2}{9c^4}\left(\ell^2 + \frac{8G^2M^2}{9c^4}\right).\label{rs4}
\end{eqnarray}
\noindent Writing $a$ and $b$ explicitly in terms of the Schwarzschild radius $R_S$ it follows from eqs.(\ref{rs3},\ref{rs4}) that
\begin{eqnarray}a&=& \frac{R_S}{3}\left(\ell^2 + \frac{R_S^2}{9}\right),\\\label{rs1}
b&=& \frac{R_S^2\ell^2}{9}\left(\ell^2 + \frac{2R_S^2}{9}\right).\label{rs2}
\end{eqnarray}
Now, substituting the values of $G$ and $c$ in the SI, and adopting $\ell \sim 0.1\, {\rm mm}$ and $M \sim 10^9 M_\odot$ (where $M_\odot \approx 2 \times 10^{33}$ g 
denotes solar mass), corresponding to 
the mass of a SMBH, it follows from eq.(\ref{111})  that the correction in the Schwarzschild radius of a SMBH 
by brane-world effects is given by 
\begin{equation}\label{coor}
R_{{\rm Sbrane}} - R_S \sim 100\,{\rm m},\end{equation}
\noindent and since the Schwarzschild radius $R_S$ is defined as $\frac{2GM}{c^2} = 2.964444 \times 10^{12}\,{\rm m}$,  the relative error 
concerning the brane-world corrections in the Schwarzschild radius of a SMBH is given by 
\begin{equation}\label{razao}
1- \frac{R_S}{R_{{\rm Sbrane}}} \sim 10^{-10}\end{equation}
\noindent
These calculations show that there exists a correction in the Schwarzschild radius of a SMBH caused by brane-world effects, although 
it is negligible. This tiny correction can be explained by the fact the event horizon of the SMBH is $10^{15}$ times bigger than the 
AdS$_5$ bulk curvature radius $\ell$. As shall be seen in a sequel paper these corrections 
are shown to be outstandingly wide in the case of mini-BHs, wherein the event horizon can be a lot of magnitude orders smaller than $\ell$.
As proved in \cite{rcp}, the solution above for $R_{{\rm Sbrane}}$ can be also found in terms of the curvature radius $\ell$.
It is then possible to find an expression for the luminosity $L$  in terms of the radius of curvature, regarding formula (\ref{dl}).

Here we shall adopt the model of the accretion rate given by a disk accretion, given by \cite{shapiro} 
Having observational values for the luminosity $L$, it is  possible to estimate a value for $\ell$, given a BH accretion model.
For a typical SMBH of $10^9 M_{\odot}$ in a massive quasar the accretion rate 
is given by 
\begin{equation}
\dot{M} \approx 2.1 \times 10^{16} {\rm kg}\, {\rm s}^{-1}
\end{equation}

Supposing the quasar radiates in Eddington limit, given by (see, e.g., \cite{shapiro})
\begin{equation}
L(\ell) = L_{{\rm Edd}} = 1.263 \times 10^{45}\left(\frac{M}{10^7 M_{\odot}}\right)\; {\rm erg\,  s^{-1}}
\end{equation}
\noindent
for a quasar with a SMBH of $10^9 M_{\odot}$, the luminosity is given by $L \sim 10^{47}\, {\rm erg\, s^{-1}}$.
From eqs.(\ref{eta}) and (\ref{dl}) the variation in quasar luminosity of a SMBH is given by 
\begin{eqnarray}\label{dell}
\Delta L &=& \frac{GM}{6c^2}\left(R_{{\rm Sbrane}}^{-1} - R_S^{-1}\right) \dot{M} c^2\nonumber\\
&=& \frac{1}{12}\left(\frac{R_S}{R_{\rm Sbrane}} -1\right)\dot{M}c^2
\end{eqnarray}\noindent 
For a typical SMBH eq.(\ref{dell}) reads
\begin{equation}\label{lop}
\Delta L \sim  10^{28}\;{\rm erg\, s^{-1}}.
\end{equation}
In terms of solar luminosity units $L_\odot = 3.9 \times 10^{33} {\rm erg\,s^{-1}}$ it follows that
the variation of luminosity of a (SMBH) quasar due to the correction of the Schwarzschild radius in a brane-world scenario is 
given by 
\begin{equation}\label{de1}
\Delta L \sim  10^{-5}\;L_\odot.
\end{equation}

Naturally, this small correction in the Schwarzschild horizon of SMBHs given by eq.(\ref{de1}) 
implies in a consequent correction in quasar luminosity via accretion mechanism. 
This correction has shown to be a hundred thousand weaker than the solar luminosity. 
In spite of the huge distance between quasars and us, it is possible these 
corrections can be never observed, although they indeed exist in a brane-world scenario. 
This correction is clearly regarded in the luminosity integrated in all wavelength. 
We look forward the detection of these corrections in particular selected wavelengths, since quasars also use to emit radiation
in the soft/hard X-ray band. In the next subsection we shall see the correction 
for a Kerr SMBH can be more notorious.

In the graphics below we illustrate the variation of luminosity $\Delta L$ of quasars as a function of the SMBH mass and $\ell$, and 
also for a given BH mass, $\Delta L$ as is depicted as a function of $\ell$. 

\begin{figure}
\includegraphics[width=8cm]{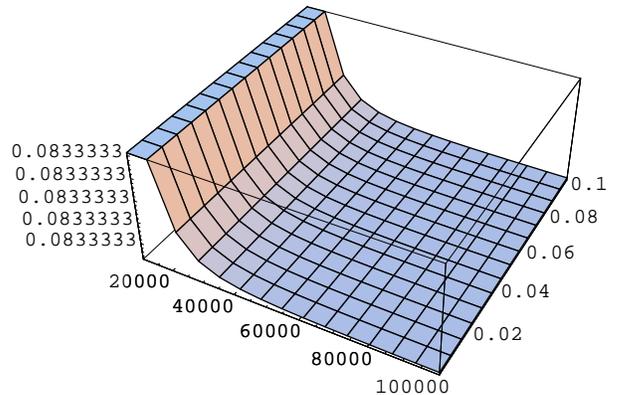}
  \caption{\small 3D graphic of $\frac{\Delta L}{\dot{M}c^2} \times \ell \times M$ where the SMBH mass $M$ varies from 10 to 10$^6$ $M_\odot$
and the radius $\ell$ of the AdS$_5$ bulk varies from $10^{-7}$ to $10^{-1}$ mm.}
\label{fig:ads51}
\end{figure}
\begin{figure}
\includegraphics[width=8cm]{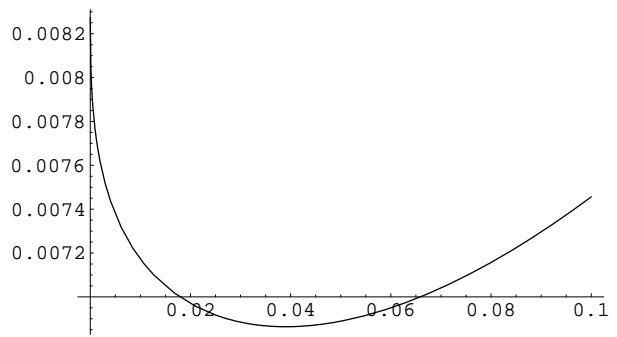}
  \caption{\small  Graphic of $\displaystyle\frac{\Delta L}{\dot{M}c^2} \times \ell$ for $M = M_\odot$}
\label{fig:ads52}
\end{figure}
\begin{figure}
\includegraphics[width=8cm]{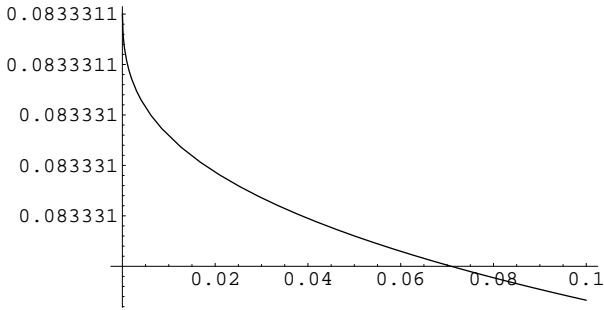}
  \caption{\small Graphic of $\displaystyle\frac{\Delta L}{\dot{M}c^2} \times \ell$ for $M = 100 M_\odot$}
\label{fig:ads53}
\end{figure}
\begin{figure}
\includegraphics[width=8cm]{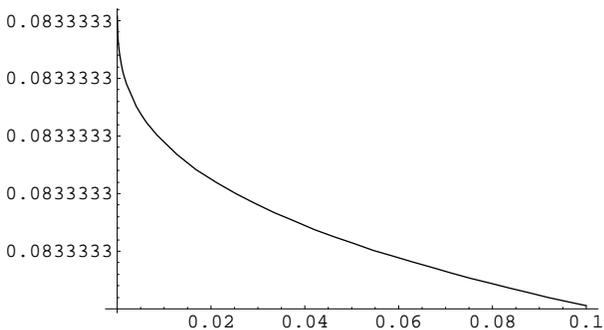}
  \caption{\small Graphic of $\displaystyle\frac{\Delta L}{\dot{M}c^2} \times \ell$ for $M = 10^4 M_\odot$}
\label{fig:ads54}
\end{figure}
{}
\subsection{Corrections in the radius of Kerr SMBHs by brane effects}
We now proceed by considering a Kerr BH on the 3-brane. 
 The Kerr metric describing the neighborhood of a spherical rotating BH with mass $M$ and angular momentum $J$, 
is given by 
\bege\label{eve} g_{\mu\nu}^{\rm Kerr} = \begin{pmatrix}
\omega^2\beta^2 - \alpha^2&0&0&-\omega\be^2\\
0&\frac{\rho^2}{\Delta}&0&0\\
0&0&\rho^2&0\\
-\omega\be^2&0&0&\be^2
\end{pmatrix}\enge\noi where \beq
\Delta &=& r^2 + \frac{a^2}{c^2} - 2\frac{GM}{c^2}r,\quad \rho^2 = r^2 + \frac{a^2}{c^2}\cos^2\theta,\label{132}\\
\Sigma^2 &=& \left(r^2 + \frac{a^2}{c^2}\right)^2 - \frac{a^2}{c^2}\Delta\sin^2\theta,\quad \be = \frac{\Sigma}{\rho}\sin\theta,\nonumber\\
\omega &=& \frac{2aGMr}{c^3\Sigma^2},\quad \alpha = \left(1 - \frac{2GM}{c^2r}\right)^{1/2}.\eeq
The rotation parameter $a$ is defined by $a = \frac{J}{Mc}$.
In order to write the Kerr metric in a diagonal form, if we solve the characteristic eigenvalue equation associated with 
eq.(\ref{eve}) the eigenvalues are given by 
\beq
\lambda_2 &=&\frac{\rho^2}{\Delta},\qquad\;
\lambda_3 = \rho^2,\nonumber\\
\lambda_{1,4} &=& \me[(\be^2 + \omega^2\beta^2 - \alpha^2) \nonumber\\
&\pm& \left((\be^2 + \omega^2\beta^2 - \alpha^2)^2 - 4((\be^2 + \omega^2\beta^2)^2 +\omega\be^2)\right)^{1/2}] \nonumber
\eeq\noi 
Now we must impose 
a condition, that arises when the eigenvalue characteristic equation is solved and we impose the condition for
\emph{real} eigenvalues, i.e., 
\bege
(1-\omega^2)^2\be^4 - 4\omega^2\be^2 + \alpha^2[4+\alpha^2 - 2\be^2(1+\omega^2)]\geq 0
\enge\noi from which the Kerr metric is given in a diagonal form: 
\beq\label{kerr}
g^{\rm Kerr} &=& g_{\mu\nu}^{\rm Kerr}dx^\mu dx^\nu\nonumber\\ 
 &=& \lambda_1 {dt^{\prime}}^2 + \frac{dr^2}{\frac{\rho^2}{\Delta}} + \rho^2 d\theta^2 + \lambda^4d{\phi^\prime}^2\label{metricap}
\eeq\noi Here $d\phi^\prime$ and $dt^\prime$ are 1-form fields on the 3-brane related to $dt$ and d$\phi$ 
by the new eigenvectors in the associated directions defined by the eigenvalue equation of eq.(\ref{eve}).

The standard internal and external Kerr radii $r_\pm$ are obtained by imposing the coefficient ${\frac{\rho^2}{\Delta}}$
  of $dr^2$ in eq.(\ref{kerr}) equals zero, i.e.,
\bege\label{raio}
R_{\pm\rm Kerr} = \frac{GM}{c^2}\pm\left(\frac{G^2M^2}{c^4} - \frac{a^2}{c^2}\right)^{1/2},
\enge\noi and for instance, for a SMBH with mass $M = 10^9 M_{\odot}$, it follows that
\beq\label{raios}
R_{+\rm Kerr} &=& 2.76586432072049\times 10^{12}\,{\rm m},\nonumber\\
 R_{-\rm Kerr} &=& 1.98580123723954\times 10^{11}\,{\rm m}.
\eeq\noi when $a = 0.5 \frac{GM}{c}$. When $a=0$, corresponding to the Schwarzschild limit (static BH) case,
it follows that $r_+ = R_S = 2.96444444444 \times 10^{12}$ m.

In order to obtain the correction of the Kerr radii, given by eq.(\ref{raio}), by brane-world effects, we follow the idea
 presented in eq.(\ref{h}). We need to find a correction that presents rotational effects, 
since the frame dragging associated with rotation of Kerr BHs
also produces deviations in the above-mentioned perturbation $\psi$, related to the Schwarzschild form in eq.(\ref{h}). In Kerr BHs, 
horizon generators are null geodesics that travel around the horizon with  angular velocity $\Omega_H$, motivating the assumption 
that the horizon itself has the rotational angular velocity $\Omega_H$. Thus, in order to incorporate rotational effects
 in eq.(\ref{psi}), we define 
 $\xi(r)$ as the deviation from the Kerr form for ${\frac{\rho^2}{\Delta}}$ (the term in the denominator of $dr^2$ in eq.(\ref{metricap})),
and, as a first trial, we suppose the \emph{ansatz}
\begin{equation}\label{133}
\xi(r) = -\frac{4GM\ell^2}{3r(c^2r^2 + a^2)}
\enge\noi in such a way that in a static limit ($a\rightarrow 0$), the correction 
in the gravitational potential satisfies eq.(\ref{potential}), and equivalently eq.(\ref{psi}).
We can prove that, at least up to rational functions with denominator polynomial of order $r^9$ and respective numerator 
polynomial of order $r^6$, 
this \emph{ansatz} is the one that best fits the limit $a\rightarrow 0$  convergence, as we explicitly show in Appendix, although 
all the  twenty six \emph{ansatzen} we investigate and analyze in Appendix 
give an  error of at most 34\%, if compared with the \emph{ansatz} given by eq.(\ref{133}). 
As we want to investigate the order of magnitude 
concerning the physical effects (as the measurability of quasar luminosity), and not the exact values, all the twenty six \emph{ansatzen}  
show the same order of magnitude associated with the variation of luminosity $\Delta L$
, for each fixed value of the rotation parameter $a$.
The more complete the \emph{ansatz} is, the more notorious the extra dimension brane-world effects are shown to be.

It can be shown that analytical functions  that are not rational, describing perturbation in the Kerr form as eq.(\ref{133}),
 do not necessarily lead Kerr coordinates to Eddington-Finkelstein coordinates in a smooth limit in its derivatives, 
in the limit when $a \rightarrow 0$. 
That is precisely the reason why we illustrate in Appendix the amount of rational functions corresponding to their 
respective \emph{ansatzen}. Indeed, it is easy to verify that when $a \rightarrow 0$ (Kerr spacetime becomes Schwarzschild), 
the Kerr coordinates become those of Eddington and Finkelstein, and the Kerr line element becomes the Eddington-Finkelstein one. 

Now, the corrections ${R_{\rm Kerrbrane}}$ in the Kerr radii, by brane-world effects, are obtained via the deviated Kerr form, as
\bege\label{1234}
 \frac{\rho^2}{\Delta} + \xi(R_{\rm Kerrbrane}) = 0.
\enge\noi By expanding the expression above and using eqs.(\ref{132}) and (\ref{133}) 
we obtain 
\beq\label{cin}
&&3c^2R_{\rm Kerrbrane}^5 - 6GMR_{\rm Kerrbrane}^4 + 6a^2R_{\rm Kerrbrane}^3 \nonumber\\&&- GM\left(6\frac{a^2}{c^2} + 4\ell^2\right)
R_{\rm Kerrbrane}^2 + \frac{3a^4}{c^2}R_{\rm Kerrbrane}\nonumber\\&& - 4GM\ell^2 \frac{a^2}{c^2}\cos^2\theta = 0.
\eeq\noi 

Using the standard values of $G$ and $c$ in the SI, and adopting again $\ell\approx 0.1$ mm and $M \sim 2 \times 10^{39}$ Kg
related to a Kerr SMBH, first 
of all the above expression has five solutions, two of which are complex conjugated each other and another one
refers to the singularity given by ${R_{\rm Kerrbrane}}\rightarrow 0$. The other two solutions correspond 
to the  (internal and external) Kerr horizons corrected by brane effects.
Eq.(\ref{cin}) has no dependence in the azimuthal angle $\theta$,  numerically at least up to 17-digit precision.
The solutions of eq.(\ref{cin}), for various values of $a$, and the relative correction  $1-\displaystyle\frac{R_{+\rm Kerr}}{R_{+\rm Kerrbrane}}$ 
in the external Kerr radius by brane effects, and consequently the variation in the quasar luminosity emission,  from  
eqs.(\ref{eta}) and (\ref{dl})  is given by 
\begin{eqnarray}\label{dell}
\Delta L &=& \frac{GM}{6c^2}\left(R_{+{\rm Kerrbrane}}^{-1}- R_{+\rm Kerr}^{-1}\right) \dot{M} c^2\nonumber\\
&=& \frac{GM}{6c^2R_{+\rm Kerr}}\left(\frac{R_{+\rm Kerr}}{R_{+\rm Kerrbrane}}-1\right)\dot{M}c^2.
\end{eqnarray}\noindent 
Here $R_{+\rm Kerrbrane}$ and $R_{-\rm Kerrbrane}$ denote respectively external and internal 
Kerr radii, corrected by brane-world extra-dimensional effects.
The results in the range $0\leq a \leq \frac{GM}{c}$ are presented in what follows:
\medbreak
\begin{center}
\begin{tabular}{||r||r|r||}\hline\hline
$\displaystyle\frac{ac}{GM}$&$1 - \displaystyle\frac{R_{+\rm Kerr}}{R_{+\rm Kerrbrane}}$&$\Delta L$ (erg s$^{-1}$)\\\hline\hline
$10^{-10}$&$\sim 10^{-7}$&$\sim 10^{28}$\\\hline
0.1&-0.00501&-7.89489 $\times 10^{31}$\\\hline
0.2&-0.02021&-3.18278 $\times 10^{32}$\\\hline
0.3&-0.04611&-7.26195 $\times 10^{32}$\\\hline
0.4&-0.08375&-1.31914 $\times 10^{33}$\\\hline
0.5&-0.13505&-2.12705 $\times 10^{33}$ \\\hline
0.6&-0.20344&-3.20417 $\times 10^{33}$ \\\hline
0.7&-0.29549&-4.65397$\times 10^{33}$ \\\hline
0.8&-0.42539&-6.69998$\times 10^{33}$\\\hline
0.9&-0.63339&-9.97583 $\times 10^{33}$ \\\hline
0.95&-0.81315&-1.28072 $\times 10^{34}$\\\hline
0.99&-1.10957&-1.74757 $\times 10^{34}$\\\hline
0.999&-1.31022&-2.06359 $\times 10^{34}$\\\hline
0.9999&-1.38048&-2.17425 $\times 10^{34}$\\\hline
0.999999&-1.41080&-2.22202 $\times 10^{34}$\\\hline\hline
\end{tabular} \end{center}\medbreak
{\small Table 1: Relative correction of Kerr radius concerning brane effects. 
$R_{+\rm Kerrbrane}$ denotes the external  brane-corrected Kerr radius.}
\medbreak
The first line in Table 1 attests the validity of the \emph{ansatz} given by eq.(\ref{133}), 
since for the limit  $a\rightarrow 0$, the Kerr radius correction by brane effects tends to the 
Schwarzschild radius correction, given by eq.(\ref{razao}). Clearly, concerning the first line 
of Table 1, the second column indeed corresponds to eq.(\ref{razao}), while the third column is related to eq.(\ref{lop}).

\begin{figure}
\includegraphics[width=8.5cm]{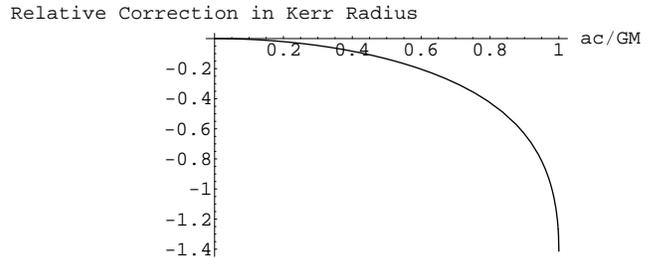}
  \caption{\small Graphic of $\left(1 - \displaystyle\frac{R_{+\rm Kerr}}{R_{+\rm Kerrbrane}}\right) \times \displaystyle\frac{ac}{GM}$
 for a SMBH.}
\label{lum3}
\end{figure}

SMBH quasar luminosity variation $\Delta L$ is given in the interval $1.27\times 10^{-2} L_\odot \leq \Delta L \leq 3.4237 L_\odot$, 
which gives for rotating (Kerr) BHs an observable correction concerning the variation in their associated quasar luminosity.
Note that $\Delta L$ is related to the external brane-corrected Kerr radius, and such corrections in $\Delta L$ are \emph{negative} ones, 
in the sense that brane effects decrease quasar observed luminosity. If we compare the results with Schwarzschild relative $\Delta L$,
the rotation of Kerr BHs are responsible for the increment of 2 up to 5 orders of magnitude in $\Delta L$.  

The results shown in Table 1 can be summarized in the graphics in Figs. (\ref{lum1}) and (\ref{lum2}).
\begin{figure}
\includegraphics[width=8.5cm]{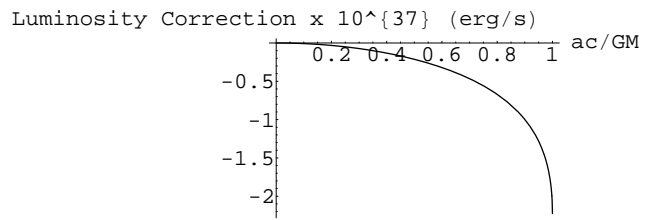}
  \caption{\small Graphic of ${\Delta L} \times \displaystyle\frac{ac}{GM}$ for $M = 10^9 M_\odot$, in the range 
$0\leq \displaystyle\frac{ac}{GM} \leq 1$.}
\label{lum1}
\end{figure}

\begin{figure}
\includegraphics[width=8.5cm]{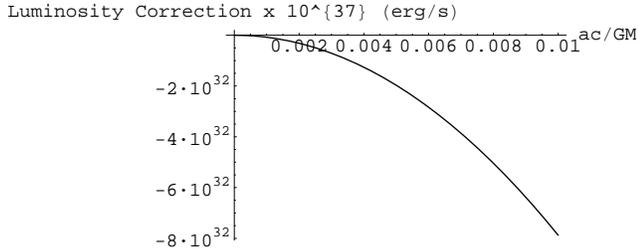}
  \caption{\small Graphic of ${\Delta L} \times \displaystyle\frac{ac}{GM}$ for $M = 10^9 M_\odot$, in the range 
$0\leq \displaystyle\frac{ac}{GM} \leq 0.01$.}
\label{lum2}
\end{figure}

\section{Concluding Remarks and Outlooks}

In the present model the variation of quasar luminosity is regarded as an extra-dimensional brane-world
 effect, and can be immediately estimated
by eq.(\ref{de1}), involving the Schwarzschild radius calculated in a brane-world scenario, and the standard 
Schwarzschild radius of a BH in spacetime.
We also investigated the more notorious brane effects regarding rotating, Kerr, BHs that
present a variation in their associated quasar luminosity of the order from $0.013$ up to  $\approx 5$ 
solar luminosities, given by eq.(\ref{dell}), also including, besides the \emph{ansatz} in eq.(\ref{133}) the other
ones given in Appendix.
These effects can be conspicuously observed, but it is hard 
to observe variation in the luminosity of Schwarzschild BHs, which we have shown 
they are of order $10^{-5} L_\odot$. Although Kerr BHs have also been investigated in the light of a brane-world
scenario \cite{stoj1,stoj2}, the correction in the Kerr internal and external radii and its consequent applications
in the observability and measurability of the variation in quasar luminosity is 
completely original, up to our knowledge. For more details on rotating black holes in brane-world scenarios with large extra dimensions see also \cite{s1,s2}.
As we show in Appendix, there are other possible \emph{ansatzen} that gives SMBH quasar luminosity variation $\Delta L$ 
in the interval $1.26\times 10^{-2} L_\odot \leq \Delta L \leq 4.3868 L_\odot$, 
which points, for rotating (Kerr) BHs, an observable correction concerning the variation in their associated quasar luminosity,
due, in particular, to the frame dragging.

Considering the fact that real quasars contain high angular momentum rotating BHs,
these results lead to an interesting viewpoint: the luminosity
observed from distant quasars are produced by bigger BHs than that
expected by nowadays previsions. An easy manner to see this is supposing
that, by observations, the luminosity of a quasar is estimated to be $L$.
Classically, a spinning black hole will produce the luminosity $L +
\Delta L$ and the corrected black hole will produce an $L$ luminosity. If
we can measure the radius of this black hole we shall see it is bigger
than the classical Kerr BH. This could shed new light in astrophysics, for instance, in  estimating the necessary time to a stellar-mass
BH to grow until be changed in a SMBH, what
naturally enlighten the formation history of galaxies in the early universe
(to see related works about the growing time of SMBHs
see, e.g., \cite{ha1,ha2,ha3}). This
idea will be developed in a forthcoming paper.

The \emph{ansatzen} used in Appendix 
lead to an  important fact: for a distant observer, when $r \gg a/c$, the \emph{ansatzen} in eq.(\ref{133}) 
and in Appendix lead to  
the static case ($\xi(r) \rightarrow \psi(r)$), and the Kerr BH can be viewed as Schwarzschild BH,  which is the same for 
$a \rightarrow 0$.
The more complete is the \emph{ansatz}, the more notorious are the effects of brane-world extra dimension in the variation
of luminosity $\Delta L$, related to quasars with Kerr BHs. 
However, variation in luminosity is caused by introduction of brane-world corrections in the BH boundaries: for a local observer, 
at the boundaries of the BH, the radius $r$
 will have magnitude comparable to $a/c$, and rotational effects will be considerable to cause corrections 
in the Kerr horizon and consequently in the quasar luminosity.

Here we want to point out it is not our main aim to accurately determine exact values for 
the variation of luminosity $\Delta L$ in quasars related to a Kerr BH, but we indeed want to emphasize
the order of magnitude of the corrections in $\Delta L$. The associated order of magnitude, for each fixed value
of the rotation parameter $a$, is always the same, and tends to increase as we put high-order polynomials
in the \emph{ansatzen}. A more complete reasoning is shown in details in Appendix.

It is also desirable 
 to calculate the variation of quasar luminosity in a Reissner-Nordstr\o m (RN) geometry generated by a charged, static BH. 
It shall be done in a sequel paper, using 
 formula, equivalent to eq.(\ref{h}) but now concerning RN metric \cite{rs06,rs09,Da}.  
 Brane models for suitable choice of the extra dimension length and of $\ell$ does predict tracks and/or
 signatures in LHC \cite{Maartens,lhc}. Black holes shall be produced in particle collisions at energies possibly below the Planck scale. 
ADD brane-worlds \cite{ad1,ad2,ad3} also provides a possibility to observe black hole production signatures in the 
next-generation colliders and cosmic ray detectors \cite{lhc1}. 
With respect to brane-world scenarios endowed with extra dimensions, 
they can solve the gauge hierarchy problem \cite{ko3,ko4,ko5,ko6} with the string scale at the
TeV scale, and the only non-supersymmetric string models
that can realize the extra-dimensional scenario, have appeared in \cite{ko1,ko2}.
 These four dimensional models are non-supersymmetric, where the $D5$-branes wrap $M^4 \times
 T^4 \times \frac{C}{\mathbb{Z}_N}$-chiral fermions get localized as open strings stretching in the
intersections between $D5$-branes and provide us with exactly the only known
string model realization of the extra-dimensional scenario of \cite{ad4}.
They  use intersecting $D5$-branes dimensional orientifold compactifications
of type IIB string theory. They also have
 two large extra dimensions transverse to the intersecting $D5$-branes and as
such one can lower the string scale to the TeV region by
keeping the Planck scale large \cite{ad4,ad2}.
Also, it must be pointed out that extra dimensions can be compactified on a curved compact hyperbolic manifold \cite{s3,s4}
 which volume grows exponentially with radius. In such model, KK spectrum is different from the flat compactification and, in particular, 
there are no light graviton modes that are the main problem for cosmology. 
Furthermore, an exponential hierarchy between the usual Planck scale and the true fundamental scale of physics 
can emerge with only order unity coefficients while the linear size of the internal space remains small. By their characteristics, 
these models are exactly in between the RS and ADD models.

In the sequel article we will show 
that, since mini-BHs possess a Reissner-Nordstr\o m-like effective behavior under gravitational potential, they 
feel a 5$D$ gravity and are more sensitive to extra dimension brane effects, and we shall repeat the procedure 
of this paper for RN mini-BHs, in order to investigate the observation of mini-BHs in LHC and the hierarchy problem.
The considerations regarding rotating ADD and RS mini-BHs \cite{s5} are to be investigated in the light of RN BHs.

\section{Acknowledgements}
The authors are grateful to Prof. Roy Maartens for his patience and clearing up expositions concerning branes.  
We are also grateful to Prof. Christos Kokorelis for bringing
into our attention some details concerning the hierarchy problem, string models and $D5$-branes,
  to Prof. Dejan Stojkovi\'c for clarifying questions regarding
rotating black holes in brane models, and to Dr. Ricardo A. Mosna for his helpful computational advisement.   
Also, the authors are grateful to \emph{JCAP} referee
for enlightening viewpoints and for the suggestions given in order to improve the quality of this paper.
CHCA thanks CNPq for financial support. 

\section{Appendix}

Besides the \emph{ansatz} given by eq.(\ref{133}), and its associated Table 1, there are other possible 
\emph{ansatzen}, among which we can list below some of them. Indeed, it 
is possible to try other functions, and their associated results are illustrated in what follows.
We present, for each \emph{ansatz} associated to $\xi(r)$, the results in their respective subsequent tables.

\begin{enumerate}

\item $
\xi(r) = -\displaystyle\frac{4GM\ell^2}{3r(c^2r^2 + car + a^2)}$
\bigskip
\begin{center}
\begin{tabular}{||r||r|r||}\hline\hline
$\displaystyle\frac{ac}{GM}$&$1 - \displaystyle\frac{R_{+\rm Kerr}}{R_{+\rm Kerrbrane}}$&$\Delta L$ (erg s$^{-1}$)\\\hline\hline
$10^{-10}$&$\sim 10^{-7}$&$\sim 10^{28}$\\\hline
0.1&-0.006012&-8.13192 $\times 10^{31}$\\\hline
0.2&-0.02464&-4.05497 $\times 10^{32}$\\\hline
0.3&-0.05109&-8.21089 $\times 10^{32}$\\\hline
0.4&-0.09645&-1.50464 $\times 10^{33}$\\\hline
0.5&-0.18128&-2.69345 $\times 10^{33}$ \\\hline
0.6&-0.25451&-4.04981 $\times 10^{33}$ \\\hline
0.7&-0.33546&-5.07349$\times 10^{33}$ \\\hline
0.8&-0.46913&-7.13873$\times 10^{33}$\\\hline
0.9&-0.69564&-1.18734 $\times 10^{34}$ \\\hline
0.95&-0.88354&-1.40993 $\times 10^{34}$\\\hline
0.99&-1.19511&-1.89631 $\times 10^{34}$\\\hline
0.999&-1.38556&-2.23880 $\times 10^{34}$\\\hline
0.9999&-1.46110&-2.37453 $\times 10^{34}$\\\hline
0.999999&-1.49347&-2.42331 $\times 10^{34}$\\\hline\hline
\end{tabular} \end{center}\medbreak
{}
\item $
\xi(r) = -\displaystyle\frac{4GM\ell^2}{3cr^2(cr + a)}$
\begin{center}
\begin{tabular}{||r||r|r||}\hline\hline
$\displaystyle\frac{ac}{GM}$&$1 - \displaystyle\frac{R_{+\rm Kerr}}{R_{+\rm Kerrbrane}}$&$\Delta L$ (erg s$^{-1}$)\\\hline\hline
$10^{-10}$&$\sim 10^{-7}$&$\sim 10^{28}$\\\hline
0.1&-0.00734&-8.87120 $\times 10^{31}$\\\hline
0.2&-0.026339&-4.08352 $\times 10^{32}$\\\hline
0.3&-0.05086&-8.07331 $\times 10^{32}$\\\hline
0.4&-0.10537&-1.46741 $\times 10^{33}$\\\hline
0.5&-0.16133&-2.71776 $\times 10^{33}$ \\\hline
0.6&-0.25613&-4.25609 $\times 10^{33}$ \\\hline
0.7&-0.34881&-5.18734$\times 10^{33}$ \\\hline
0.8&-0.43524&-7.04734$\times 10^{33}$\\\hline
0.9&-0.69335&-1.06498 $\times 10^{34}$ \\\hline
0.95&-0.86209&-1.38912 $\times 10^{34}$\\\hline
0.99&-1.15977&-1.88530 $\times 10^{34}$\\\hline
0.999&-1.36129&-2.24378 $\times 10^{34}$\\\hline
0.9999&-1.44376&-2.36108 $\times 10^{34}$\\\hline
0.999999&-1.46312&-2.41906 $\times 10^{34}$\\\hline\hline
\end{tabular} \end{center}\medbreak

\item $
\xi(r) = -\displaystyle\frac{4GM\ell^2r}{3(c^2r^2 + a^2)^2}
$
\begin{center}
\begin{tabular}{||r||r|r||}\hline\hline
$\displaystyle\frac{ac}{GM}$&$1 - \displaystyle\frac{R_{+\rm Kerr}}{R_{+\rm Kerrbrane}}$&$\Delta L$ (erg s$^{-1}$)\\\hline\hline
$10^{-10}$&$\sim 10^{-7}$&$\sim 10^{28}$\\\hline
0.1&-0.00913&-8.54973 $\times 10^{31}$\\\hline
0.2&-0.02956&-4.15322 $\times 10^{32}$\\\hline
0.3&-0.05731&-8.64175 $\times 10^{32}$\\\hline
0.4&-0.10142&-1.56713 $\times 10^{33}$\\\hline
0.5&-0.21965&-2.89342 $\times 10^{33}$ \\\hline
0.6&-0.29231&-4.28744 $\times 10^{33}$ \\\hline
0.7&-0.38300&-5.18456$\times 10^{33}$ \\\hline
0.8&-0.51241&-7.39810$\times 10^{33}$\\\hline
0.9&-0.73087&-1.23884 $\times 10^{34}$ \\\hline
0.95&-0.92635&-1.45102 $\times 10^{34}$\\\hline
0.99&-1.27391&-1.97662 $\times 10^{34}$\\\hline
0.999&-1.57328&-2.38209 $\times 10^{34}$\\\hline
0.9999&-1.61298&-2.63901 $\times 10^{34}$\\\hline
0.999999&-1.50139&-2.59463 $\times 10^{34}$\\\hline\hline
\end{tabular} \end{center}\medbreak

\item $
\xi(r) = -\displaystyle\frac{4GM\ell^2r}{3(c^2r^2 + a^2)(c^2r^2 + car + a^2)}
$
\begin{center}
\begin{tabular}{||r||r|r||}\hline\hline
$\displaystyle\frac{ac}{GM}$&$1 - \displaystyle\frac{R_{+\rm Kerr}}{R_{+\rm Kerrbrane}}$&$\Delta L$ (erg s$^{-1}$)\\\hline\hline
$10^{-10}$&$\sim 10^{-7}$&$\sim 10^{28}$\\\hline
0.1&-0.00967&-8.67821 $\times 10^{31}$\\\hline
0.2&-0.03174&-4.21755 $\times 10^{32}$\\\hline
0.3&-0.059334&-8.72392 $\times 10^{32}$\\\hline
0.4&-0.12890&-1.59335 $\times 10^{33}$\\\hline
0.5&-0.24509&-2.94252 $\times 10^{33}$ \\\hline
0.6&-0.31823&-4.33756 $\times 10^{33}$ \\\hline
0.7&-0.41002&-5.23908$\times 10^{33}$ \\\hline
0.8&-0.53771&-7.43905$\times 10^{33}$\\\hline
0.9&-0.77346&-1.26771 $\times 10^{34}$ \\\hline
0.95&-0.95327&-1.49214 $\times 10^{34}$\\\hline
0.99&-1.30631&-2.01136 $\times 10^{34}$\\\hline
0.999&-1.60455&-2.41774 $\times 10^{34}$\\\hline
0.9999&-1.64224&-2.66329 $\times 10^{34}$\\\hline
0.999999&-1.54598&-2.64902 $\times 10^{34}$\\\hline\hline
\end{tabular} \end{center}\medbreak

\item $
\xi(r) = -\displaystyle\frac{4GM\ell^2(r+\frac{a}{c})}{3r^2(c^2r^2 + a^2)}
$
\begin{center}
\begin{tabular}{||r||r|r||}\hline\hline
$\displaystyle\frac{ac}{GM}$&$1 - \displaystyle\frac{R_{+\rm Kerr}}{R_{+\rm Kerrbrane}}$&$\Delta L$ (erg s$^{-1}$)\\\hline\hline
$10^{-10}$&$\sim 10^{-7}$&$\sim 10^{28}$\\\hline
0.1&-0.00821&-8.64319 $\times 10^{31}$\\\hline
0.2&-0.02235&-3.94566 $\times 10^{32}$\\\hline
0.3&-0.04862&-7.98111 $\times 10^{32}$\\\hline
0.4&-0.09134&-1.41398 $\times 10^{33}$\\\hline
0.5&-0.15423&-2.63452 $\times 10^{33}$ \\\hline
0.6&-0.22957&-3.95534 $\times 10^{33}$ \\\hline
0.7&-0.30414&-5.00451$\times 10^{33}$ \\\hline
0.8&-0.43524&-6.97134$\times 10^{33}$\\\hline
0.9&-0.66819&-1.02457 $\times 10^{34}$ \\\hline
0.95&-0.83415&-1.35673 $\times 10^{34}$\\\hline
0.99&-1.13776&-1.84235 $\times 10^{34}$\\\hline
0.999&-1.33411&-2.17996 $\times 10^{34}$\\\hline
0.9999&-1.41389&-2.32723 $\times 10^{34}$\\\hline
0.999999&-1.43313&-2.37076 $\times 10^{34}$\\\hline\hline
\end{tabular} \end{center}\medbreak
{}
\item $
\xi(r) = -\displaystyle\frac{4GM\ell^2r}{(c^2r^2 + car + a^2)^2}
$
\begin{center}
\begin{tabular}{||r||r|r||}\hline\hline
$\displaystyle\frac{ac}{GM}$&$1 - \displaystyle\frac{R_{+\rm Kerr}}{R_{+\rm Kerrbrane}}$&$\Delta L$ (erg s$^{-1}$)\\\hline\hline
$10^{-10}$&$\sim 10^{-7}$&$\sim 10^{28}$\\\hline
0.1&-0.00842&-8.91420 $\times 10^{31}$\\\hline
0.2&-0.02529&-4.02583 $\times 10^{32}$\\\hline
0.3&-0.05182&-8.03184 $\times 10^{32}$\\\hline
0.4&-0.09372&-1.46297 $\times 10^{33}$\\\hline
0.5&-0.15814&-2.68207 $\times 10^{33}$ \\\hline
0.6&-0.25996&-4.04617 $\times 10^{33}$ \\\hline
0.7&-0.34845&-5.10553$\times 10^{33}$ \\\hline
0.8&-0.47972&-7.17852$\times 10^{33}$\\\hline
0.9&-0.70341&-1.07640 $\times 10^{34}$ \\\hline
0.95&-0.89352&-1.42865 $\times 10^{34}$\\\hline
0.99&-1.21774&-1.95393 $\times 10^{34}$\\\hline
0.999&-1.39224&-2.26408 $\times 10^{34}$\\\hline
0.9999&-1.48285&-2.43071 $\times 10^{34}$\\\hline
0.999999&-1.52619&-2.66318 $\times 10^{34}$\\\hline\hline
\end{tabular} \end{center}\medbreak

\item $
\xi(r) = -\displaystyle\frac{4GM\ell^2(r + \frac{a}{c})}{(c^2r^2 + a^2)^2}
$
\begin{center}
\begin{tabular}{||r||r|r||}\hline\hline
$\displaystyle\frac{ac}{GM}$&$1 - \displaystyle\frac{R_{+\rm Kerr}}{R_{+\rm Kerrbrane}}$&$\Delta L$ (erg s$^{-1}$)\\\hline\hline
$10^{-10}$&$\sim 10^{-7}$&$\sim 10^{28}$\\\hline
0.1&-0.00953&-8.98053 $\times 10^{31}$\\\hline
0.2&-0.02741&-4.07342 $\times 10^{32}$\\\hline
0.3&-0.05671&-8.09554 $\times 10^{32}$\\\hline
0.4&-0.09912&-1.50712 $\times 10^{33}$\\\hline
0.5&-0.19623&-2.75923 $\times 10^{33}$ \\\hline
0.6&-0.30745&-4.23076 $\times 10^{33}$ \\\hline
0.7&-0.39610&-5.30651$\times 10^{33}$ \\\hline
0.8&-0.52886&-7.38309$\times 10^{33}$\\\hline
0.9&-0.76124&-1.36297 $\times 10^{34}$ \\\hline
0.95&-0.96100&-1.74522 $\times 10^{34}$\\\hline
0.99&-1.29054&-2.29364 $\times 10^{34}$\\\hline
0.999&-1.47301&-2.51908 $\times 10^{34}$\\\hline
0.9999&-1.56772&-2.76221 $\times 10^{34}$\\\hline
0.999999&-1.60182&-2.84310 $\times 10^{34}$\\\hline\hline
\end{tabular} \end{center}\medbreak

\item $
\xi(r) = -\displaystyle\frac{4GM\ell^2(r + \frac{a}{c})}{(c^2r^2 + car +  a^2)^2}
$
\begin{center}
\begin{tabular}{||r||r|r||}\hline\hline
$\displaystyle\frac{ac}{GM}$&$1 - \displaystyle\frac{R_{+\rm Kerr}}{R_{+\rm Kerrbrane}}$&$\Delta L$ (erg s$^{-1}$)\\\hline\hline
$10^{-10}$&$\sim 10^{-7}$&$\sim 10^{28}$\\\hline
0.1&-0.00908&-8.94376 $\times 10^{31}$\\\hline
0.2&-0.02595&-4.05387 $\times 10^{32}$\\\hline
0.3&-0.054621&-8.06410 $\times 10^{32}$\\\hline
0.4&-0.09741&-1.48326 $\times 10^{33}$\\\hline
0.5&-0.18845&-2.72164 $\times 10^{33}$ \\\hline
0.6&-0.28644&-4.20988 $\times 10^{33}$ \\\hline
0.7&-0.37232&-5.27359$\times 10^{33}$ \\\hline
0.8&-0.49871&-7.34237$\times 10^{33}$\\\hline
0.9&-0.73420&-1.31489 $\times 10^{34}$ \\\hline
0.95&-0.92755&-1.70367 $\times 10^{34}$\\\hline
0.99&-1.24309&-2.22785 $\times 10^{34}$\\\hline
0.999&-1.40463&-2.44762 $\times 10^{34}$\\\hline
0.9999&-1.50478&-2.66423 $\times 10^{34}$\\\hline
0.999999&-1.53287&-2.68250 $\times 10^{34}$\\\hline\hline
\end{tabular} \end{center}\medbreak

\item $
\xi(r) = -\displaystyle\frac{4GM\ell^2(r^2+ \frac{ar}{c} + \frac{a^2}{c^2})}{3r^3(c^2r^2 + car + a^2)}
$
\begin{center}
\begin{tabular}{||r||r|r||}\hline\hline
$\displaystyle\frac{ac}{GM}$&$1 - \displaystyle\frac{R_{+\rm Kerr}}{R_{+\rm Kerrbrane}}$&$\Delta L$ (erg s$^{-1}$)\\\hline\hline
$10^{-10}$&$\sim 10^{-7}$&$\sim 10^{28}$\\\hline
0.1&-0.00878&-8.91935 $\times 10^{31}$\\\hline
0.2&-0.02415&-3.98723 $\times 10^{32}$\\\hline
0.3&-0.05064&-8.01176 $\times 10^{32}$\\\hline
0.4&-0.18546&-1.43512 $\times 10^{33}$\\\hline
0.5&-0.19754&-2.84356 $\times 10^{33}$ \\\hline
0.6&-0.239758&-4.01574 $\times 10^{33}$ \\\hline
0.7&-0.31145&-5.12985$\times 10^{33}$ \\\hline
0.8&-0.44786&-7.34130$\times 10^{33}$\\\hline
0.9&-0.68111&-1.05427 $\times 10^{34}$ \\\hline
0.95&-0.86921&-1.39712 $\times 10^{34}$\\\hline
0.99&-1.25475&-1.96459 $\times 10^{34}$\\\hline
0.999&-1.38733&-2.31902 $\times 10^{34}$\\\hline
0.9999&-1.45616&-2.42564 $\times 10^{34}$\\\hline
0.999999&-1.46692&-2.43704 $\times 10^{34}$\\\hline\hline
\end{tabular} \end{center}\medbreak
{}
\item $\xi(r) = -\displaystyle\frac{4GM\ell^2}{3c^2r^3}\exp\left(\frac{ac}{GM}\right)$
\begin{center}
\begin{tabular}{||r||r|r||}\hline\hline
$\displaystyle\frac{ac}{GM}$&$1 - \displaystyle\frac{R_{+\rm Kerr}}{R_{+\rm Kerrbrane}}$&$\Delta L$ (erg s$^{-1}$)\\\hline\hline
$10^{-10}$&$\sim 10^{-7}$&$\sim 10^{28}$\\\hline
0.1&-0.0178&-9.0935 $\times 10^{31}$\\\hline
0.2&-0.0295&-4.1017 $\times 10^{32}$\\\hline
0.3&-0.0528&-8.1352 $\times 10^{32}$\\\hline
0.4&-0.2031&-1.4645 $\times 10^{33}$\\\hline
0.5&-0.2165&-2.9081 $\times 10^{33}$ \\\hline
0.6&-0.2546&-4.1024 $\times 10^{33}$ \\\hline
0.7&-0.3345&-5.2371$\times 10^{33}$ \\\hline
0.8&-0.4657&-7.4152$\times 10^{33}$\\\hline
0.9&-0.7018&-1.1276 $\times 10^{34}$ \\\hline
0.95&-0.8901&-1.4464 $\times 10^{34}$\\\hline
0.99&-1.2879&-2.0183 $\times 10^{34}$\\\hline
0.999&-1.412&-2.4132 $\times 10^{34}$\\\hline
0.9999&-1.493&-2.4739 $\times 10^{34}$\\\hline
0.999999&-1.502&-2.4916 $\times 10^{34}$\\\hline\hline
\end{tabular} \end{center}\medbreak

\item $
\xi(r) = -\displaystyle\frac{4GM\ell^2(r^3+ \frac{a^2r}{c^2} + \frac{ar^2}{c})}{3r^4(c^2r^2 + car + a^2)}
$
\begin{center}
\begin{tabular}{||r||r|r||}\hline\hline
$\displaystyle\frac{ac}{GM}$&$1 - \displaystyle\frac{R_{+\rm Kerr}}{R_{+\rm Kerrbrane}}$&$\Delta L$ (erg s$^{-1}$)\\\hline\hline
$10^{-10}$&$\sim 10^{-7}$&$\sim 10^{28}$\\\hline
0.1&-0.0095&-8.9876 $\times 10^{31}$\\\hline
0.2&-0.0270&-4.1746 $\times 10^{32}$\\\hline
0.3&-0.0545&-8.0823 $\times 10^{32}$\\\hline
0.4&-0.2098&-1.4734 $\times 10^{33}$\\\hline
0.5&-0.2154&-2.9065 $\times 10^{33}$ \\\hline
0.6&-0.2543&-4.0854 $\times 10^{33}$ \\\hline
0.7&-0.3390&-5.2108$\times 10^{33}$ \\\hline
0.8&-0.4654&-7.4832$\times 10^{33}$\\\hline
0.9&-0.7179&-1.0924 $\times 10^{34}$ \\\hline
0.95&-0.8951&-1.4132 $\times 10^{34}$\\\hline
0.99&-1.2873&-1.9946 $\times 10^{34}$\\\hline
0.999&-1.4190&-2.3397 $\times 10^{34}$\\\hline
0.9999&-1.4815&-2.4452 $\times 10^{34}$\\\hline
0.999999&-1.5185&-2.5304 $\times 10^{34}$\\\hline\hline
\end{tabular} \end{center}\medbreak

Here we want to point out it is not our main aim to accurately determine exact values for 
the variation of luminosity in quasars related to a Kerr BH, but in this first article, we only want to emphasize
the order of magnitude of this correction in $\Delta L$. The associated order of magnitude, for each fixed value
of the rotation parameter $a$, is always the same, including other \emph{ansatzen}, like the rational functions listed below:
\beq
\xi(r) &=& -\displaystyle\frac{4GM\ell^2r^2}{3(c^3r^3 + a^3)(c^2r^2 + a^2)},\nonumber\\
 \xi(r) &=& 
-\displaystyle\frac{4GM\ell^2r^2}{3(c^3r^3 + a^3)(c^2r^2 + car + a^2)}\nonumber\\
\xi(r) &=& -\displaystyle\frac{4GM\ell^2r^2}{3(c^3r^3 + c^2r^2a + a^3)(c^2r^2 + car + a^2)},\nonumber\\
\xi(r) &=& -\displaystyle\frac{4GM\ell^2r^2}{3(c^3r^3 + c^2ar^2 + ca^2r + a^3)(c^2r^2 + car + a^2)}\nonumber\eeq\beq
\xi(r) &=& -\displaystyle\frac{4GM\ell^2(r^2 + \frac{a^2}{c^2})}{3(c^3r^3 + a^3)(c^2r^2 + a^2)},\nonumber\\ \xi(r) &=& 
-\displaystyle\frac{4GM\ell^2(r^2+\frac{a^2}{c^2})}{3(c^3r^3 + a^3)(c^2r^2 + car + a^2)}\nonumber\\
\xi(r) &=& -\displaystyle\frac{4GM\ell^2(r^2+\frac{a^2}{c^2})}{3(c^3r^3 + c^2r^2a + a^3)(c^2r^2 + car + a^2)},\nonumber\\
\xi(r) &=& -\displaystyle\frac{4GM\ell^2(r^2+\frac{a^2}{c^2})}{3(c^3r^3 + c^2ar^2 + ca^2r + a^3)(c^2r^2 + car + a^2)},\nonumber\\
\xi(r) &=& -\displaystyle\frac{4GM\ell^2(r^2+car+\frac{a^2}{c^2})}{3(c^3r^3 + c^2ar^2 + ca^2r + a^3)(c^2r^2 + car + a^2)},\nonumber\\
\xi(r) &=& -\displaystyle\frac{4GM\ell^2(r^2+car+\frac{a^2}{c^2})}{3(c^3r^3 + c^2ar^2  + a^3)(c^2r^2 + car + a^2)},\nonumber\\
\xi(r) &=& -\displaystyle\frac{4GM\ell^2(r^2+car+\frac{a^2}{c^2})}{3(c^3r^3 + c^2ar^2 + ca^2r)(c^2r^2 + car + a^2)},\nonumber\\
\xi(r) &=& -\displaystyle\frac{4GM\ell^2(r^3+\frac{a}{c}r^2+\frac{a^2}{c^2})}{3(c^3r^3 + c^2ar^2 + ca^2r + a^3)^2},\nonumber\\
\xi(r) &=& -\displaystyle\frac{4GM\ell^2(r^3+\frac{a}{c}r^2+\frac{a^2}{c^2})}{3(c^3r^3 + c^2ar^2 + ca^2r + a^3)(c^3r^3 + c^2ar^2 + a^3)},\nonumber\\
\xi(r) &=& -\displaystyle\frac{4GM\ell^2(r^3+\frac{a}{c}r^2+\frac{a^2}{c^2})}{3(c^3r^3 + c^2ar^2 + ca^2r + a^3)(c^3r^3 + ca^2r + a^3)},\nonumber\\
\xi(r) &=& -\displaystyle\frac{4GM\ell^2(r^3+\frac{a}{c}r^2+\frac{a^2}{c^2})}{3(c^3r^3  + ca^2r + a^3)(c^3r^3 + c^2ar^2 + a^3)},\nonumber
\eeq\noi for instance.
\end{enumerate}

All those formul\ae\, for the \emph{ansatz} related to $\xi(r)$ give a value for $\Delta L$ which is shown to be at most 34\% 
bigger than the original \emph{ansatz} given by eq.(\ref{133}). Anyway, the RS brane-world extra-dimensional effects, 
concerning the variation in quasar luminosity $\Delta L$, are always
more significant than the original deviation given by eq.(\ref{1234}), coming from eq.(\ref{133}).

\end{document}